\documentclass[aps,pra,twocolumn,amssymb,superscriptaddress]{revtex4-1}

\usepackage{graphicx}
\usepackage{amsmath}
\usepackage{verbatim}
\usepackage{physics}
\usepackage{hyperref}

\begin{document}
\title{Feshbach resonances in potassium Bose-Bose mixtures}
\author{L. Tanzi}
\altaffiliation[Present address: ] {CNR-INO, S.S. \textquotedblleft A. Gozzini\textquotedblright\,   di Pisa, via Moruzzi 1, 56124 Pisa, and LENS and Dipartimento di Fisica e Astronomia, Universit\`{a} di Firenze, 50019 Sesto Fiorentino, Italy}
\affiliation{ICFO - Institut de Ci\`{e}ncies Fot\`{o}niques, The Barcelona Institute of Science and Technology, 08860 Castelldefels (Barcelona), Spain}
\author{C. R. Cabrera}
\affiliation{ICFO - Institut de Ci\`{e}ncies Fot\`{o}niques, The Barcelona Institute of Science and Technology, 08860
Castelldefels (Barcelona), Spain}
\author{J. Sanz}
\affiliation{ICFO - Institut de Ci\`{e}ncies Fot\`{o}niques, The Barcelona Institute of Science and Technology, 08860
Castelldefels (Barcelona), Spain}
\author{P. Cheiney}
\altaffiliation[Present address: ] {iXblue, 34 rue de la Croix de Fer, 78105 Saint-Germain-en-Laye, France}
\affiliation{ICFO - Institut de Ci\`{e}ncies Fot\`{o}niques, The Barcelona Institute of Science and Technology, 08860
Castelldefels (Barcelona), Spain}
\author{M. Tomza}
\affiliation{Faculty of Physics,  University of Warsaw, Pasteura 5, 02-093 Warsaw, Poland}
\author{L. Tarruell}
\email[Electronic address: ]{leticia.tarruell@icfo.eu}
\affiliation{ICFO - Institut de Ci\`{e}ncies Fot\`{o}niques, The Barcelona Institute of Science and Technology, 08860
Castelldefels (Barcelona), Spain}

\date{\today}
\begin{abstract}
We present a detailed study of the scattering properties of ultracold mixtures of bosonic potassium atoms. We locate 20 previously unobserved Feshbach resonances in isotopic $^{39}$K--$^{41}$K mixtures. These are assigned to $s$-wave molecular channels by comparison to an asymptotic bound state model and coupled channels calculations. Additional Feshbach resonances are studied in spin mixtures of a single potassium isotope, both in $^{39}$K and $^{41}$K. In particular, we characterize the parameters of a selected $^{39}$K Feshbach resonance by radio-frequency association of Feshbach molecules. Our results could be exploited to refine the model potentials for potassium scattering. Furthermore, these new Feshbach resonances enlarge the range of experiments possible with degenerate Bose-Bose mixtures.
\end{abstract}

\maketitle

\section{Introduction}

Magnetic Feshbach resonances provide a powerful tool for controlling the interatomic interactions in ultracold atomic systems \cite{ChinRMP2010}. Over the last 20 years, they have enabled the study of a wealth of few- and many-body phenomena. Key examples include the magneto-association of ultracold Feshbach molecules \cite{JulienneRMP2006}, the observation of Efimov bound states \cite{FerlainoFBS2011} and the study of strongly-interacting systems, a prime example of which are unitary limited gases \cite{Zwerger2011,ChevyRev2016}.

Interacting quantum mixtures, obtained either by cooling and trapping different species or by combining different internal states of the same atom, give access to a particularly rich playground. From the few-body point of view, extensive experimental efforts have enabled the association of heteronuclear mixtures into ground-state diatomic molecules, paving the way towards the realization of long-range interacting quantum systems \cite{BohnRev2017, DeMarco2018}. From the many-body perspective, quantum mixtures of ultracold atoms have been exploited to realize prototypical condensed matter models, giving access to new phenomena or extreme parameter regimes not accessible in \textquotedblleft natural\textquotedblright\, systems. The observation of phase separation in repulsive Bose-Bose and Bose-Fermi mixtures \cite{HallPRL1998,PappPRL2008,NicklasPRL2011,GrimmPRL2018}, the study of strongly interacting Fermi and Bose polarons \cite {MassignanRPP2014,Jin2016,JoergensenPRL2016}, the realization of superfluid Bose-Fermi mixtures \cite{FerrierScience2014}, the observation of ultradilute quantum liquid droplets in attractive Bose-Bose mixtures \cite{PetrovPRL2015, CabreraScience2018,CheineyPRL2018,SemeghiniPRL2018}, and the realization of systems with mediated interactions \cite{DeSalvoArXiv2018} constitute some landmark examples.

A good overlap between the different components of the mixture is crucial in most of these studies. This can be easily achieved by employing atoms with identical (or very similar) masses and polarizabilities, which therefore experience identical external potentials. Quantum mixtures comprised of different internal states of the same atom or isotopic combinations of the same chemical element naturally fulfill this requirement. Controlling the interactions in such systems requires the presence of suitable Feshbach resonances. Interstate resonances have been demonstrated in $^{87}$Rb \cite{MartePRL2002,WideraPRL2004,ErhardPRA2004} and $^{39}$K \cite{JoergensenPRL2016}, whereas isotopic resonances have been explored in $^{85}$Rb--$^{87}$Rb \cite{PappPRL2006}, $^6$Li--$^7$Li \cite{ZhangICAP2005} and $^{40}$K--$^{41}$K \cite{WuPRA2011} mixtures.

In this paper, we explore the Feshbach resonances available in Bose-Bose mixtures of potassium atoms. We perform a systematic study of the isotopic $^{39}$K--$^{41}$K mixture, locating 20 previously unobserved Feshbach resonances. Their positions are determined through atom loss spectroscopy, and a consistent assignment to molecular levels is carried out using an asymptotic bound state model (ABM) \cite{WillePRL2008,TieckePRA2010}. Coupled channels (CC) calculations are then employed to perform a full analysis of the width and position of the resonances using the model potentials for potassium scattering proposed in Ref. \cite{FalkePRA2008}. We explore as well Feshbach resonances in mixtures of internal states of a single potassium isotope. For $^{41}$K we locate a new interstate Feshbach resonance, and for $^{39}$K we perform radio-frequency spectroscopy of the molecular state of a selected resonance that is particularly convenient from the experimental point of view. Combining experimental observations, CC calculations and an analytical model, we provide a precise characterization of its properties, including its range parameter. Our results could be used to refine the existing potassium model potentials \cite{DErricoNJP2007,FalkePRA2008,PashovEPJD2008}, and to look for a possible breakdown of the Born-Oppenheimer approximation -- and thus of the mass scaling normally used to predict isotopic collisions. Furthermore, the resonances characterized in this work could be exploited in the future to explore different classes of many-body phenomena in multi-component bosonic systems.

This article is organized as follows. In Section \ref{sec:Exp}, we briefly describe the experimental procedures. Section \ref{sec:39-41} is devoted to the scattering properties of $^{39}$K--$^{41}$K mixtures and describes the ABM and CC calculations employed in the rest of the paper. Section \ref{sec:41} focuses on the Feshbach resonances of $^{41}$K available at magnetic fields $\sim50$~G. In Section \ref{sec:39} we study a selected interstate Feshbach resonance of $^{39}$K and develop a simple theoretical model to determine the parameters characterizing it. Finally, in Section \ref{sec:Conclusion} we draw the conclusions of this work.

\section{Experimental preparation\label{sec:Exp}}

In order to perform Feshbach spectroscopy of all possible potassium Bose-Bose mixtures, we prepare ultracold clouds in a crossed optical dipole trap and subject them to a homogeneous magnetic field in the range of $50-650$ G. Our experimental sequence starts with a double-isotope 3D MOT of $^{41}$K and $^{39}$K, which is sequentially loaded from a 2D$^{+}$ MOT. After a sub-Doppler cooling stage, consisting of blue-detuned D1 grey molasses for $^{41}$K
\cite{SieversPRA2015, ChenPRA2016}, and red-detuned D2 molasses for $^{39}$K \cite{LandiniPRA2011}, both isotopes are optically pumped into the $|F=2,m_F=2\rangle$ state and loaded into a quadrupole magnetic trap. Here $F$ is the total angular momentum and $m_F$ its projection along the magnetic field direction. Forced evaporative cooling of $^{41}$K is performed on the hyperfine transition, whereas $^{39}$K is cooled sympathetically. The atoms are then transferred into a hybrid trap created by the addition of a far-detuned optical dipole trap beam ($\lambda=1064$ nm, $w\simeq65\,\mu$m) positioned $\sim60\,\mu$m below the zero of the quadrupole magnetic field \cite{LinPRA2009}. After a second stage of evaporation, the cloud is loaded into a crossed beam optical dipole trap with trap frequencies $\omega/2\pi=\left[67(5),163(5),176(5)\right]$ Hz. All experiments reported in this paper are performed in similar conditions, with $\sim 10^5$ atoms of both isotopes and a temperature of $\sim 420$ nK. This corresponds to thermal clouds with $T/T_c\simeq10$, where $T_c$ denotes the critical temperature for Bose-Einstein condensation. For experiments involving only $^{41}$K we suppress the $^{39}$K MOT loading sequence. Pure samples of $^{39}$K are obtained by eliminating $^{41}$K after the hybrid trap evaporation using a short light pulse resonant with the $F=2\rightarrow F'=3$ transition.

\section{$^{39}$K--$^{41}$K mixtures\label{sec:39-41}}

\begin{table*}[t!]
\begin{ruledtabular}
\begin{tabular}{lrrrrrrr}
Entrance channel                                & $M_F$     & $B_0^\mathrm{exp}$ [G] & $B_0^\mathrm{ABM}$ [G] & $\delta\mu\,[\mu_{\mathrm{B}}]$ & $B_0^\mathrm{CC}$ [G]    & $\Delta^\mathrm{CC}$ [G] & $a_\mathrm{bg}^\mathrm{CC} [a_0]$ \\
\hline
$^{39}$K$|1,1\rangle$+$^{41}$K$|2,2\rangle$     & $3\,\,\,$ & 341.5(2)                  & 340.608                     & 1.56                      & 341.619                   & 0.138                     & 135.2 \\
                                                & $3\,\,\,$ & 353.8(3)                  & 351.706                     & 1.26                      & 354.010                   & 0.493                     & 135.2 \\
\hline

$^{39}$K$|1,1\rangle$+$^{41}$K$|1,1\rangle$     & $2\,\,\,$ & 139.27(4)                  & 139.122                     & -2.97                     & 139.400                   & 0.0374                    & 173.0 \\
                                                & $2\,\,\,$ & 146.24(7)               & 146.011                     & -2.45                     & 146.411                   & 0.111                     & 173.0 \\
                                                & $2\,\,\,$ & 338.12(7)                  & 337.758                     & -1.95                     & 338.281                   & 0.0461                    & 176.4 \\
                                                & $2\,\,\,$ & 500.2(3)                  & 495.592                     & -0.73                     & 500.049                   & 0.700                     & 176.1 \\
                                                & $2\,\,\,$ & 518.4(1)                  & 516.038                     & -1.56                     & 518.433                   & 0.128                     & 176.1 \\
\hline
$^{39}$K$|1,1\rangle$+$^{41}$K$|1,0\rangle$     & $1\,\,\,$ &  88.2(1)                  & 68.898                      & 0.02                      &  88.475                   & 0.0258                    & 168.4 \\
                                                & $1\,\,\,$ & 160.05(6)                  & 159.805                     & -2.62                     & 160.128                   & 0.0474                    & 172.5 \\
                                                & $1\,\,\,$ & 165.80(5)                  & 165.409                     & -2.23                     & 165.933                   & 0.110                     & 172.5 \\
                                                & $1\,\,\,$ & 344.4(1)                  & 343.864                     & -1.93                     & 344.509                   & 0.128                     & 176.2 \\
                                                & $1\,\,\,$ & 522.6(2)                  & 518.198                     & -0.78                     & 522.478                   & 0.621                     & 176.0 \\
                                                & $1\,\,\,$ & 553.1(1)                  & 550.218                     & -1.30                     & 552.964                   & 0.198                     & 176.0 \\

\hline
$^{39}$K$|1,1\rangle$+$^{41}$K$|1,-1\rangle$    & $0\,\,\,$ & 189.88(5)                  & 189.343                     & -2.97                     & 189.999                   & 0.0766                    & 172.6 \\
                                                & $0\,\,\,$ & 348.4(1)                  & 347.567                     & -1.92                     & 348.463                   & 0.180                     & 176.2 \\
                                                & $0\,\,\,$ & 384.91(7)                  & 384.631                     & -1.92                     & 385.073                   & 0.0635                    & 175.4 \\
                                                & $0\,\,\,$ & 553.5(2)                  & 549.759                     & -0.94                     & 553.378                   & 0.506                     & 176.1 \\
\hline
$^{39}$K$|1,0\rangle$+$^{41}$K$|1,-1\rangle$    & $-1\,\,\,$& 228.88(8)                  & 228.256                     & -0.88                     & 229.039                   & 0.989                     & 171.8 \\

\hline
$^{39}$K$|1,-1\rangle$+$^{41}$K$|1,-1\rangle$   & $-2\,\,\,$& 149.84(6)                  & 145.561                     & 0.07                      & 149.764                   &-0.0252                    & 163.7 \\

\hline
$^{39}$K$|1,-1\rangle$+$^{41}$K$|2,-2\rangle$   & $-3\,\,\,$& 649.6(6)                  & 645.937                     & -1.00                     & 649.167                   & 0.673                     & 176.2 \\
\end{tabular}
\end{ruledtabular}
\caption{Summary of the Feshbach resonances between $^{39}$K and $^{41}$K observed in this work. The experimental positions $B_0^\mathrm{exp}$ are determined from Gaussian fits to the measured loss features, and the reported uncertainties correspond to the $1/\mathrm{e}^2$ half-width of the Gaussian fit. $B_0^\mathrm{ABM}$ and $\delta\mu$ denote the theoretical predictions of the asymptotic bound state model (ABM) for the resonance position and the difference in magnetic moments for the open and closed channels respectively. $B_0^\mathrm{CC}$, $\Delta^\mathrm{CC}$ and $a_\mathrm{bg}^\mathrm{CC}$ are the results of a coupled channels (CC) calculation using the model potentials of Ref. \cite{FalkePRA2008}. The CC systematic uncertainty is $\sim0.3$ G and the number of significant digits given only indicates the precision of the calculation.} \label{Table1}
\end{table*}

\begin{figure*}[t]
\includegraphics[width=2\columnwidth,clip=true]{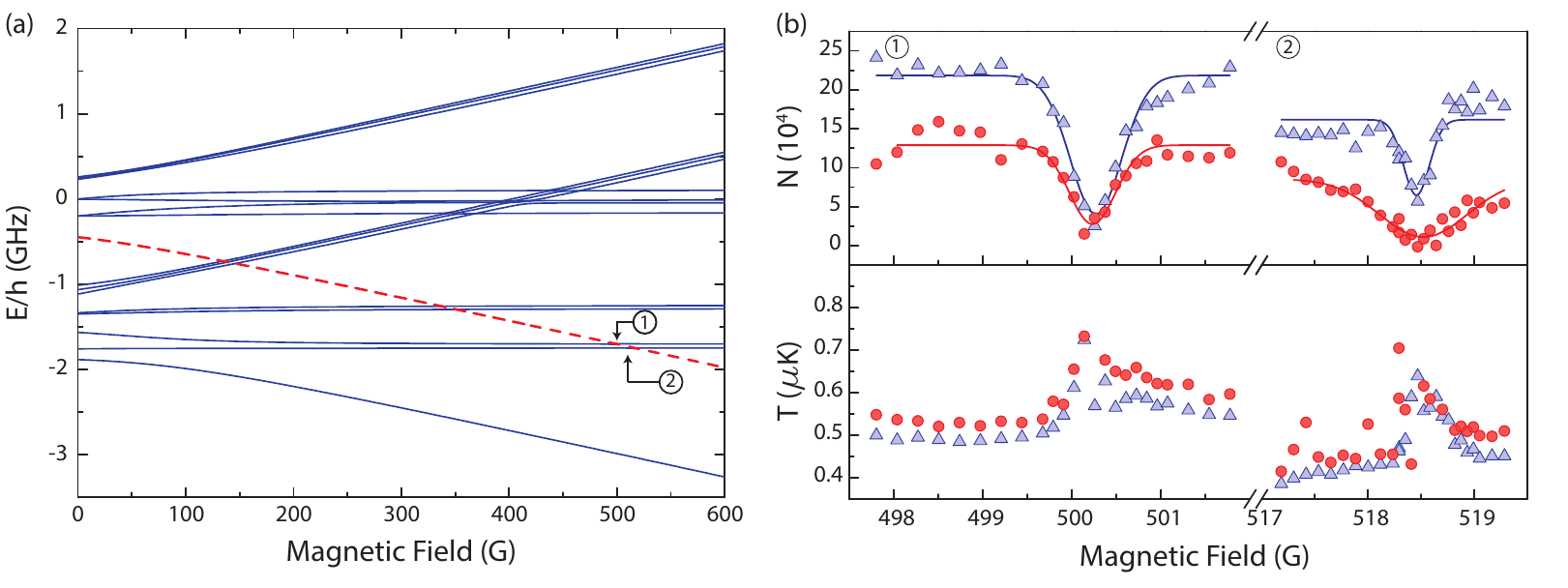}
\caption{Feshbach resonances for
$^{39}$K$|1,1\rangle$+$^{41}$K$|1,1\rangle$ collisions. (a) Asymptotic bound state model (ABM) molecular state energies. Blue solid lines: open-channel threshold energies. Red dashed line: uncoupled $s$-wave molecular state energy. (b) Experimental atom loss and temperature spectra for the $^{39}$K$|1,1\rangle$ (red circles) and $^{41}$K$|1,1\rangle$ (blue triangles) states. The solid line is the empirical Gaussian fit used to extract the resonance positions. Two interisotope Feshbach resonances are observed at $\sim500$ G, in agreement with the ABM predictions.} \label{Fig1}
\end{figure*}

In the first series of experiments we locate Feshbach resonances in the $^{39}$K--$^{41}$K mixture. This atomic combination  was not studied experimentally before, and provides the largest mass difference between potassium isotopes. It could thus be used for verifying the validity of the Born-Oppenheimer approximation used in Ref. \cite{FalkePRA2008} to derive interisotope model interaction potentials based on $^{39}$K--$^{39}$K spectroscopic data. From a many-body physics perspective, the resonances reported below are particularly suitable for the study of stable three-component bosonic systems with a good spatial overlap between them. For instance, they would enable the realization of magnetic polarons: quantum impurities immersed in a magnetic background comprised of a two-component Bose-Einstein condensate and dressed not only by phonon but also by magnon excitations \cite{AshidaPRB2018}. In the $^{39}$K--$^{41}$K system, this situation could be achieved by identifying two miscible states of $^{41}$K with the medium and one state of $^{39}$K with the impurity.

In order to provide sufficient data for a complete characterization of the interisotope interaction potentials, we explore scattering channels corresponding to all $M_F=m_F^{39}+m_F^{41}$ values where the mixture is in the lowest energy state and does not experience inelastic spin-exchange collisions. The different spin state combinations are prepared using Landau-Zener radio-frequency sweeps. Feshbach spectroscopy is then performed by ramping the magnetic field to the desired value and measuring simultaneously the atom number of $^{39}$K and $^{41}$K after a variable hold time, which is adjusted empirically in order to optimize the experimental signal. A resonance manifests itself as an enhancement of atomic losses due to an increase of the inelastic three-body recombination rate, which can be observed as a decrease in the atom number of both isotopes and an increase of their temperature.
We find a total of 20 loss features within the range $88-650$ G, which are summarized in Table \ref{Table1}. Given the small width of the observed features (between $10$ mG and $100$ mG), we identify the position of each resonance with the center of the corresponding loss curve. Note that this could lead to systematic errors in the determined positions, with a bias towards the positive scattering length side of the resonances \cite{BourdelPRL2003}. For all the measurements the magnetic field is calibrated to a precision of $10$ mG using radio-frequency transitions between Zeeman sublevels in the vicinity of each of the loss features.

To perform a first assignment of the observed Feshbach resonances and identify the molecular channels responsible for them, we have adapted the asymptotic bound state model (ABM) introduced in Ref. \cite{WillePRL2008} to the $^{39}$K--$^{41}$K mixture. The ABM takes as input parameters the energies of the last \emph{s-}wave bound states of the singlet and triplet interatomic potentials. Their coupling depends on the overlap between the corresponding molecular wave-functions, which are additional parameters of the model. We obtain the energies and Franck-Condon factors from the interaction potentials of Ref. \cite{FalkePRA2008}. In order to reproduce all of the observed loss features, the two last vibrational states of both potentials need to be taken into account. The corresponding energies and Franck-Condon factors are $E_S^1/h= -32.1(2)$ MHz, $E_T^1/h = -8.33(5)$ MHz, $E_S^2/h= -1698.1(2.2)$ MHz, $E_T^2/h =-1282.5(9)$ MHz, $\eta_{11}=0.9180$, $\eta_{22}=0.9674$, $\eta_{12}=0.0895$ and $\eta_{21}=0.0463$. Here $S (T)$ denotes a singlet (triplet) bound state, $vv'$ (with $v(v')=1,2$) the overlap between the singlet and triplet wavefunctions in the vibrational state $v$ and $v'$ respectively, and $h$ is Planck's constant.
Fig. \ref{Fig1}(a) displays the molecular state energies predicted by the ABM for the absolute ground state of the system $^{39}$K$|1,1\rangle$+$^{41}$K$|1,1\rangle$. Fig. \ref{Fig1}(b) shows typical loss features and the corresponding temperature increase measured simultaneously for the two isotopes, from which the resonance positions are extracted.

A more accurate description of the scattering properties of the investigated systems is obtained by performing a coupled channels (CC) calculation using the same model potentials as above \cite{FalkePRA2008}. We restrict our calculations to $s$-wave collisions and $s$-wave resonances only, which is a justified and accurate approximation in the absence of overlapping $s$-wave and higher partial-wave resonances. We neglect the spin-dipole/spin-dipole interaction and second-order spin-orbit coupling responsible for the  dipolar relaxation, which are expected to have negligible effect on the $s$-wave resonances.
The coupled channels equations are solved as in Refs.~\cite{TomzaPRL14,TomzaPRA15a}, assuming a temperature of $100$ nK in all calculations. The resonance position $B_0$, the resonance width $\Delta$, and the local background scattering length $a_\mathrm{bg}$ are obtained by fitting the numerical points with the analytical expression
\begin{equation}
a(B)=a_\mathrm{bg}+a_\mathrm{res}=a_\mathrm{bg}\left(1-\frac{\Delta}{B-B_0}\right).
\label{eq:a}
\end{equation}
When two \emph{s}-wave resonances overlap, we use instead the more general expression
\cite{LangePRA2009}
\begin{equation}
a(B)=a_\mathrm{bg}\left(1-\frac{\Delta_1}{B-B_{0,1}}\right)\left(1-\frac{\Delta_2}{B-B_{0,2}}\right)\label{overlappinga},
\end{equation}
where the subscripts denote each of the resonances.

The results of the CC calculation are summarized in Table \ref{Table1}. We quantify the agreement with the experimental data through $\delta=B_0^\mathrm{exp}-B_0^\mathrm{CC}$, which yields a rms deviation for all resonances of $0.24$ G. This is on the order of the average error bars of the experimental results and of the systematic uncertainty of the CC model ($\sim0.3$~G, due to the uncertainty of the potentials and associated singlet and triplet scattering lengths). The CC calculations are performed with mass scaled model potentials. Although the predictions seem accurate enough in view of the experimental uncertainties, a more detailed analysis of the current measurements could be exploited to analyze the validity of the Born-Oppenheimer approximation used to derive the model potentials \cite{FalkePRA2008, PashovEPJD2008}.

\section{$^{41}$K spin mixture\label{sec:41}}

In a second series of experiments we study the scattering properties of $^{41}$K spin mixtures. We locate a previously unobserved Feshbach resonance in the $|1,0\rangle+|1,-1\rangle$ channel  around $50$ G, in the vicinity of the single-component resonance experimentally studied in Ref. \cite{KishimotoPRA2009}. This situation provides good control over both the interstate and intrastate scattering lengths, and makes $^{41}$K an interesting system for the study of two-component Bose gases with repulsive intrastate and tunable interstate interactions. In the attractive case, quantum droplets in a system with symmetric intrastate scattering lengths could be created \cite{PetrovPRL2015}. In the presence of a coherent coupling between the two states, the repulsive case is well suited to study the transition from a paramagnetic to a ferromagnetic phase \cite{RecatiVarenna2016}. If the coupling is created using two-photon transitions, systems combining tunable interactions and synthetic gauge fields could be explored \cite{GoldmanRoPP2014}.

\begin{figure}[t]
\includegraphics[width=1\columnwidth,clip=true]{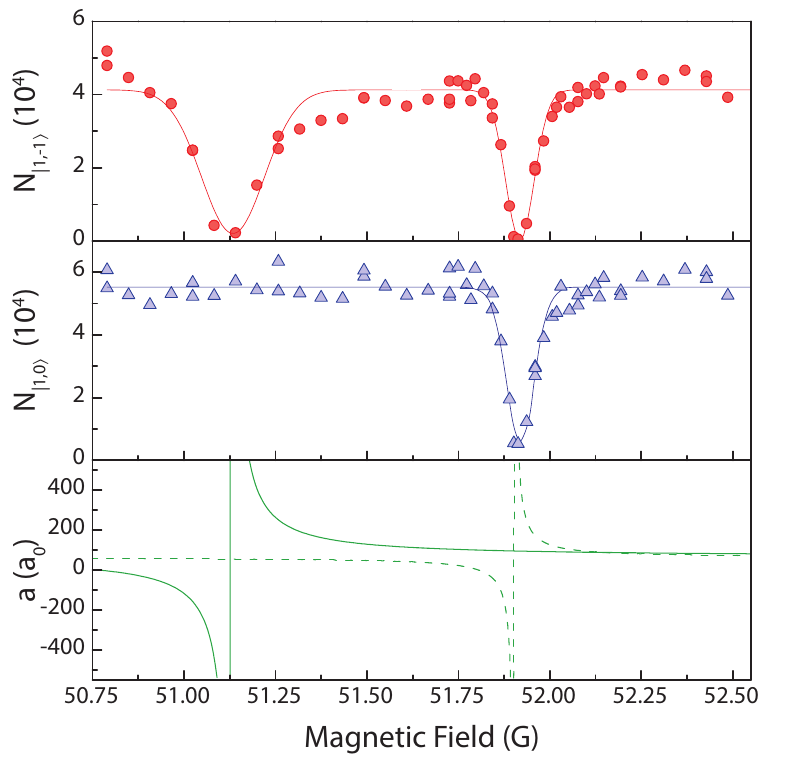}
\caption{Feshbach resonances for $^{41}$K collisions in the $|1,-1\rangle +|1,-1\rangle$ and $|1,0\rangle +|1,-1\rangle$ scattering channels. Top and middle panels: experimentally observed loss features for the $|1,-1\rangle$ (red circles) and $|1,0\rangle$ (blue triangles) states. The solid line is the empirical Gaussian fit used to extract the resonance positions. Bottom panel: scattering lengths for $|1,-1\rangle +|1,-1\rangle$ (continuous line) and $|1,-1\rangle +|1,0\rangle$ (dashed line) collisions predicted by our CC calculations. The $|1,0\rangle +|1,0\rangle$ scattering length does not vary in this magnetic field range, and remains close to the background value $\sim65a_0$.}
\label{Fig2}
\end{figure}

Fig. \ref{Fig2} displays a loss spectroscopy measurement performed in both the $|1,0\rangle$ and $|1,-1\rangle$ channels, together with the relevant interstate and intrastate scattering lengths predicted by our CC model. As before, we employ the mass scaled model potentials of Ref. \cite{FalkePRA2008}. Experimentally, we locate the position of the two Feshbach resonances from Gaussian fits to the measured loss features. The $|1,-1\rangle +|1,-1\rangle$ resonance is found at $B_0=51.1(2)$ G, in agreement with the value $B_0=51.129$ G predicted by the CC calculation and previous experimental and theoretical works \cite{KishimotoPRA2009, DErricoNJP2007, LyseboPRA2010}. The theoretical width of this resonance is $\Delta=-0.361$ G, with a local background scattering length $a_{\mathrm{bg}}/a_0=65.1$.

Atom losses in both the $|1,0\rangle$ and $|1,-1\rangle$ states signal the presence of an interstate Feshbach resonance at $B_0=51.92(8)$ G. Although this feature was not observed experimentally before, it was theoretically predicted to occur at a magnetic field $B_0=51.95$ G \cite{LyseboPRA2010}. This is in good agreement with our measurement. Our experimental findings are consistent with our CC model, which predicts a position $B_0=51.903$ G, a resonance width $\Delta=-0.0978$ G and a local background scattering length $a_{\mathrm{bg}}/a_0=65.1$.

\section{$^{39}$K spin mixture\label{sec:39}}
We perform the last series of experiments in a spin mixture of $^{39}$K. This isotope has several single-component Feshbach resonances \cite{DErricoNJP2007,RoyPRL2013} that have been exploited for studying a broad range of phenomena including disorder physics \cite{RoatiNature2008, DErricoPRL2014}, unitary Bose gases \cite{ChevyRev2016}, matter-wave bright solitons \cite{LepoutrePRA2016}, and quantum liquid droplets \cite{PetrovPRL2015, CabreraScience2018, CheineyPRL2018,SemeghiniPRL2018}. Recently, a resonance in the $|1,0\rangle +|1,-1\rangle$ channel with a large width $\Delta\simeq 16$ G was reported. Since it is in a magnetic field range where the scattering length of state $|1,-1\rangle$ is approximately constant and positive, it is very well adapted to the experimental study of strongly-interacting Bose polarons using this state as a bath and state $|1,0\rangle$ as an impurity \cite{JoergensenPRL2016}. Here we perform a more accurate determination of its properties by measuring the binding energy of the Feshbach molecules. Combined with our CC model, this allows us to obtain accurate values of the resonance position, width, background scattering length, range parameter and strength, as we detail in the following.

\begin{figure}[t]
\includegraphics[width=1\columnwidth,clip=true]{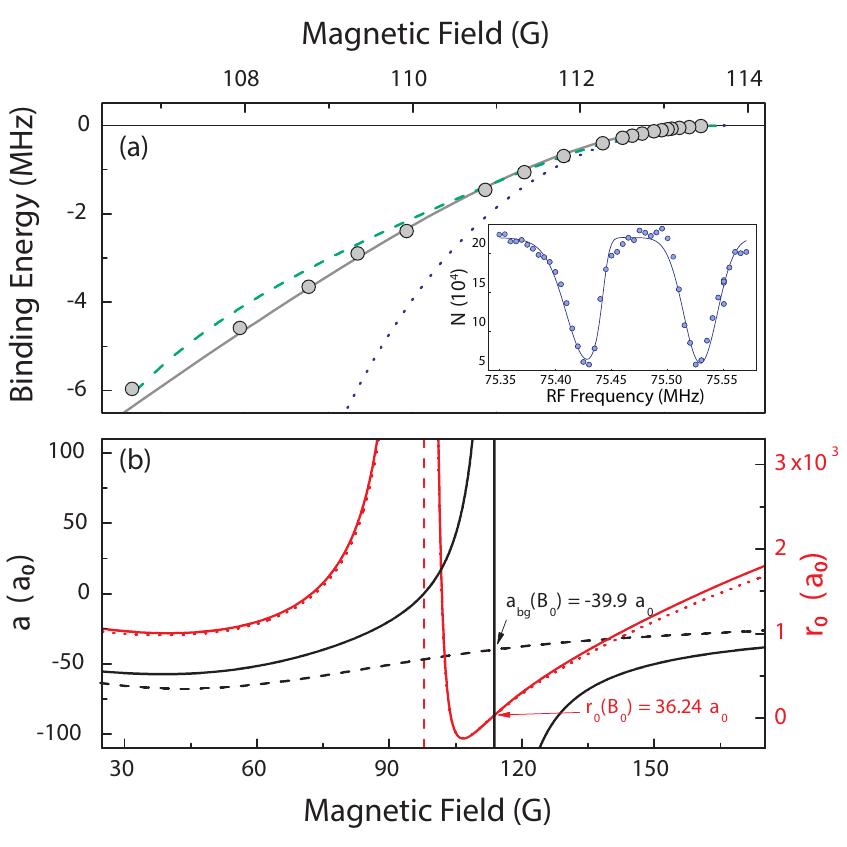}
\caption{Characterization of an interstate Feshbach resonance in the $^{39}$K $|1,0\rangle +|1,-1\rangle$ scattering channel. (a) Molecular binding energy as a function of the magnetic field. Grey dots: experimental measurements. The error bars ($\sim10$ kHz) are smaller than the size of the symbol. Blue dotted line: universal formula. Grey solid line: CC prediction. Green dashed line: low energy expansion of the binding energy Eq. (\ref{eq:Eb}), extracted from the CC scattering length and effective range. Inset: typical radio-frequency association spectrum. The solid line is a fit to the model used to extract $E_{\mathrm{b}}$ from the distance between the molecular (left) and atomic (right) peaks. (b) Scattering length $a$ and effective range $r_0$ predicted by the CC model (black and red solid lines, respectively). $r_0$ can also be computed combining Eqs. (\ref{eq:r01}) and (\ref{eq:r02}) (red dotted line). The red dashed line represents the magnetic field $B_{\mathrm{zc}}$ where $r_0$ diverges and the scattering length crosses zero. The black dashed line represents the background scattering length $a_{\mathrm{bg}}(B)$, as explained in the main text.}
\label{fig:MolecularSpectroscopy}
\end{figure}

\paragraph*{Experimental results.}
Two different techniques are employed to accurately measure the molecular binding energy $E_{\mathrm{b}}$. In the close vicinity of the resonance, we directly associate molecules starting from a $|1,0\rangle +|1,-1\rangle$ mixture and subjecting it to a modulated magnetic field of frequency corresponding to $E_{\mathrm{b}}$. For larger binding energies, we exploit instead radio-frequency association of Feshbach molecules. Starting from the non-resonant state $|1,0\rangle$, we apply a radio-frequency pulse close to the $|1,0\rangle$ to $|1,-1\rangle$ transition and transfer atoms to the bound state. In both cases, the formation of molecules is signaled by a reduction of the trapped atom number. Indeed, the molecules decay due to vibrational quenching induced by collisions with unpaired atoms and escape from the trap. The inset of Fig. \ref{fig:MolecularSpectroscopy}(a) displays a typical radio-frequency spectrum. The association of molecules corresponds to the asymmetric feature, which reflects the fact that the association frequency depends on the kinetic energy of the atom pair forming the molecule. We model it by the convolution of a Maxwell-Boltzmann distribution and a Gaussian function (solid line) \cite{WeberPRA2008}. The latter results from the finite molecule lifetime and from technical broadening. From the onset of the molecular feature we obtain $E_{\mathrm{b}}$. The radio-frequency spectrum displays an additional peak, corresponding to the atomic hyperfine transition, which is absent in the data obtained \emph{via} magnetic field modulation.

Fig. \ref{fig:MolecularSpectroscopy}(a) summarizes the measured molecular binding energy as a function of the magnetic field. Near the resonance, it follows the universal relation $E_\mathrm{b}=-\hbar^2/m a^2$, where $m$ denotes the mass of $^{39}$K. At larger detunings, finite range corrections become important and this simple expression loses its validity. In order to provide a model-independent parametrization of the scattering length we thus restrict ourselves to the range $B-B_0\le 1$ G, where the non-universal corrections are negligible \footnote{We empirically determine this range by fitting the experimental data to the universal formula in increasing magnetic field ranges, and evaluating the quality of each fit \emph{via} a $\chi^2$ test. We observe that the extracted resonance parameters and the $\chi^2$ value remain nearly unchanged as long as $B-B_0\le 1$ G, corresponding to the 10 smallest experimental values of $E_{\mathrm{b}}$ shown in Fig. \ref{fig:MolecularSpectroscopy}(a).}. The validity of this assumption will be discussed below. The fit to the universal formula (blue dotted line) yields $B_0=113.76(1)$ G for the resonance position, and $a_{\mathrm{bg}}\Delta/a_0=715(7)$ G for the product of the resonance width and local background scattering length. The error bars represent the fit uncertainty, whereas the systematic uncertainty on our magnetic field calibration is on the order of 10 mG.

\paragraph*{Theoretical model.}
We exploit the CC model to predict the complete low temperature scattering properties of the gas in the vicinity of this resonance. Aside from the molecular binding energy $E_{\mathrm{b}}$ and the scattering length $a$, we also compute the effective range $r_0$. It corresponds to the first finite-momentum correction to the scattering amplitude and can be viewed as an energy dependence of the scattering length,
\begin{equation}
-\frac{1}{a(\epsilon)}=-\frac{1}{a}+\frac{1}{2}r_0 k^2+...
\end{equation}
Here, $\epsilon=\hbar^2 k^2 /m$ is the energy in the center of mass frame. We obtain $r_0$ by fitting the numerical points calculated for six collision energies between 100 nK and 400 nK to this expression.

The CC results are summarized in Fig. \ref{fig:MolecularSpectroscopy}. In particular, the grey solid line in panel (a) represents $E_{\mathrm{b}}$, while the black and red solid lines in panel (b) represent $a$ and $r_0$, respectively. Unlike the universal fit, which holds only in the vicinity of the resonance pole, the CC model provides a good description of the molecular binding energy measurements in the complete magnetic field range. The resonance position $B_0$ agrees well with our experimental determination. However, given the uncertainty of the CC calculations ($\sim0.3$ G), the experimental results give a more reliable value. By fitting the scattering length $a$ close to the resonance center with Eq. (\ref{eq:a}), we obtain $a_{\mathrm{bg}}\Delta/a_0=727(6)$ G. Here, the error bar represents the uncertainty of the fitted parameters. This value is consistent with the one extracted from the universal fit to the experimental data, validating the CC model. The CC calculation also provides the position of the zero crossing of the scattering length $B_{\mathrm{zc}}=97.85$ G. There, the effective range $r_0$ diverges.

The computed scattering length and effective range can be combined to predict analytically the binding energy away from the universal regime \cite{DykePRA2013}
\begin{equation}
E_{\mathrm{b}}=-\frac{\hbar^2}{m r_0^2}\left(\sqrt{1-2r_0/a}-1 \right)^2 \label{eq:Eb}.
\end{equation}
This low energy expansion is valid for $a\gg 2 r_0$. It coincides with the universal expression only for large values of the scattering length ($a\gg 2R_{\mathrm{vdW}}$, where $R_{\mathrm{vdW}}/a_0=64.61$ denotes the van der Waals length \cite{FalkePRA2008}). The latter condition is fulfilled for $B=(112.8-113.75)$ G, validating \emph{a posteriori} the range used to determine experimentally the resonance parameters. In Fig. \ref{fig:MolecularSpectroscopy}(a) the low energy expansion Eq. (\ref{eq:Eb}) is depicted as the green dashed line. In its regime of validity it agrees with both the experimental points and the CC calculations. This confirms the validity of the CC prediction for the effective range.

Besides the resonance feature, Fig. \ref{fig:MolecularSpectroscopy}(b) shows a strong magnetic field dependence of the scattering length far from the resonance pole. This complicates the estimation of the standard resonance parameters. In the following, we provide a simple model characterizing the magnetic field dependence of both $a$ and $r_0$ in order to determine accurate values for the local background scattering length, width, range parameter, and strength of this Feshbach resonance.

Close to the resonance, the scattering length $a(B)=a_{\mathrm{bg}}(B)+a_{\mathrm{res}}(B)$ is dominated by the resonant contribution $a_{\mathrm{res}}(B)=a_{\mathrm{bg}}(B_0)\Delta/(B_0-B)$, where $a_{\mathrm{bg}}(B_0)$ is the background scattering length on resonance. The value of the background scattering length varies considerably in the relevant magnetic field range. This is due to the change in singlet and triplet admixture of the states involved as a function of $B$, and to the large difference between the singlet and triplet scattering lengths ($a_S/a_0= 138.80$ and $a_T/a_0=−33.41$). We empirically determine the function $a_{\mathrm{bg}}(B)$ by subtracting from $a(B)$ the resonant contribution, and fitting the result with a polynomial function \footnote{The magnetic field dependence of the background scattering length is well described by the empirical expression $a_{\mathrm{bg}}/a_0=-53.4672-0.173989 B-0.0244375 B^2+0.000769723 B^3-8.27353\times 10^{-6} B^4+4.16737\times 10^{-8} B^5-9.78912\times 10^{-11} B^6+8.19883\times 10^{-14} B^7$, where the magnetic field $B$ is expressed in G.}. As shown in Fig. \ref{fig:MolecularSpectroscopy}(b) (black dashed line), $a_{\mathrm{bg}}(B)$ varies asymmetrically by $\sim 40 a_0$ across the resonance position. On resonance, the background scattering length is $a_{\mathrm{bg}}(B_0)/a_0=-39.9$, which implies $\Delta=-18.2$ G.
Note that, due to the asymmetric dependence of the scattering length with magnetic field below and above $B_0$, identifying the width $\Delta$ with the difference $B_0-B_{\mathrm{zc}}\simeq 15.9$ G  is misleading for this resonance. For this reason, our determination of $\Delta$ differs considerably from the one of Ref. \cite{JoergensenPRL2016}, $\Delta=-15.88$ G, although the product $a_{\mathrm{bg}}\Delta/a_0=720$ G reported there is in good agreement with our findings.

Similarly, the effective range can be divided into a background and a resonant contribution $r_0(B)=r_{0,\mathrm{bg}}(B)+r_{0,\mathrm{res}}(B)$.
For a van der Waals potential, the non-resonant part depends on the magnetic field according to \cite{Flambaum93, Gao98, ChinRMP2010}
\begin{equation}
r_{0,\mathrm{bg}}(B)=\frac{\Gamma(1/4)^4}{6\pi^2}\bar{a}\left[1-2\frac{\bar{a}}{a(B)}+2\left(\frac{\bar{a}}{a(B)}\right)^2\right],
\label{eq:r01}
\end{equation}
where $\bar{a}=4\pi/\Gamma(1/4)^2 R_{\mathrm{vdW}}$.
On the other hand, the resonant contribution is given by \cite{WernerEPJB2009,MassignanEPL2012}
\begin{equation}
r_{0,\mathrm{res}}(B)=-2 R_*\left(1-\frac{a_{\mathrm{bg}}(B)}{a(B)}\right)^2,
\label{eq:r02}
\end{equation}
where $R_*$ is the so-called range parameter \cite{PetrovPRL2004}. Following Ref. \cite{RoyPRL2013}, the range parameter can be obtained by combining Eqs. (\ref{eq:r01}) and (\ref{eq:r02}), and evaluating them at $B=B_0$. Explicitly,
\begin{equation}
 R_*=-\frac{r_0(B_0)}{2}+\frac{\Gamma(1/4)^2 R_{\mathrm{vdW}}}{3\pi}.
\end{equation}
Using the numerical value $r_0(B_0)/a_0=36.24$ we find $R_*/a_0=71.99$. The red dotted line in Fig. \ref{fig:MolecularSpectroscopy}(b) represents $r_0$, computed as the sum of Eqs. (\ref{eq:r01}) and (\ref{eq:r02}). Remarkably, these simple analytical expressions are in good agreement with the CC results (red solid line) provided the magnetic field dependence of the background scattering length is taken into account.

Alternatively, we can evaluate the range parameter using the expression $R_*=\hbar^2/m a_{\mathrm{bg}}\Delta\delta\mu$ \cite{PetrovPRL2004}, which yields $R_*/a_0\simeq 77$. There, we use the relative magnetic moment $\delta \mu/\mu_{\mathrm{B}}\simeq1.18$ from the ABM prediction and the CC value for $a_{\mathrm{bg}}\Delta$. Extracting the range parameter from the effective range is expected to be more precise because for $^{39}$K the relative magnetic moment changes with magnetic field due to broad avoided crossings between near-threshold molecular states \cite{DErricoNJP2007, RoyPRL2013}. Both results differ considerably from the value $R_*/a_0=60 a_0$ proposed in \cite{JoergensenPRL2016}. From the range parameter, we deduce the strength parameter of the resonance \cite{ChinRMP2010} $s_{\mathrm{res}}=\bar{a}/R_*= 0.86$. This value corresponds to an intermediate coupling regime, indicating that despite its large magnetic field width the resonance is not open channel dominated.

\section{Conclusions\label{sec:Conclusion}}
In conclusion, we have performed a comprehensive experimental and theoretical study of the low energy scattering properties of various potassium Bose-Bose mixtures. We have located 20 previously unobserved Feshbach resonances in the mixture $^{39}$K--$^{41}$K, and confirmed the validity of the model potentials proposed in Ref. \cite{FalkePRA2008} for this isotopic combination. At the level of accuracy of experiment and theory, this constitutes a test of the validity of the Born-Oppenheimer approximation for potassium scattering. For $^{41}$K we have identified a new interstate Feshbach resonance that gives access to two-component Bose gases with repulsive intrastate interactions and tunable interstate ones. For $^{39}$K we have measured the binding energy of the molecular state responsible for the interstate Feshbach resonance at $\sim114$ G \cite{JoergensenPRL2016}. Combining experimental observations, CC calculations and analytical models, we have provided a precise characterization of the resonance parameters, including the resonance strength and its range parameter. The model employed to analyze this specific resonance incorporates explicit calculations of the field-dependent effective range and background scattering length and accounts for non-universal finite-range effects. Due to its simplicity, this model could be used to characterize other Feshbach resonances.

 The 23 Feshbach resonances characterized in this work could be exploited to improve the model potentials for potassium scattering. Such efforts should include as well the new $^{40}$K \cite{LudewigPhD2012,KrauserPRA2017}, $^{41}$K \cite{ChenPRA2016} and $^{40}$K--$^{41}$K resonances \cite{WuPRA2011} that have been located during the last years, and the most recent measurements on single-component $^{39}$K gases \cite{RoyPRL2013,FletcherScience2017}. Furthermore, the tunability of scattering lengths available in potassium Bose-Bose mixtures opens a wealth of possibilities for future research. The $^{39}$K--$^{41}$K mixture enables the realization of three-component Bose gases with tunable interactions, where magnetic polarons have been predicted \cite{AshidaPRB2018}. In the presence of coherent coupling, the $^{41}$K interstate resonance is well suited for the study of the paramagnetic-ferromagnetic phase transition \cite{RecatiVarenna2016}, and for adding interaction control to experiments with synthetic gauge fields in synthetic dimensions \cite{CeliPRL2014,ManciniScience2015,StuhlScience2015}. Finally, the $^{39}$K interstate resonance constitutes an ideal system for the observation of Bose polarons in the strongly interacting regime \cite{JoergensenPRL2016}.
\\

\acknowledgments
We thank M.~Bosch-Aguilera, V.~Lienhard, and J.~Sastre for experimental assistance during the construction of the experimental apparatus, C. Chisholm and A. Fr\"{o}lian for a careful reading of the manuscript, and J.~C.~Cifuentes, C.~Dengra, X.~Menino and the corresponding ICFO facilities for technical support. We acknowledge A. Simoni for insightful discussions and for giving us the results of Refs. \cite{DErricoNJP2007, RoyPRL2013} in numerical form. The experimental work was financially supported by Fundaci\'{o} Privada Cellex, European Union (MagQUPT-631633 and QUIC-641122), Ministerio de Ciencia, Innovaci\'{o}n y Universidades (StrongQSIM FIS2014-59546-P, QuDROP FIS2017-88334-P and Severo Ochoa SEV-2015-0522), Deutsche Forschungsgemeinschaft (FOR2414), and Generalitat de Catalunya (SGR1660 and CERCA program). C.R.C. acknowledges support from the Consejo Nacional de Ciencia y Tecnolog\'{\i}a (402242/ 384738), J.S. from the Ministerio de Ciencia, Innovaci\'{o}n y Universidades (BES-2015-072186), P.C. from the Marie Sk{\l}odowska-Curie actions (TOPDOL-657439), M.T. from the National Science Centre Poland (2016/23/B/ST4/03231) and the PL-Grid Infrastructure, and L. Tarruell from the Ministerio de Ciencia, Innovaci\'{o}n y Universidades through the Ram\'{o}n y Cajal program (RYC-2015-17890).

\end{document}